\documentclass[journal]{IEEEtran}
\usepackage{amsmath}

\ifCLASSINFOpdf
\else
\fi

\usepackage{graphicx} 
\usepackage{booktabs}

\begin{document}
%
\title{Hybrid Quantum CNN for Cross-Sensor Spaceborne Volcanic Thermal Activity Recognition Worldwide }
%
%
%

\author{Claudia~Corradino,
Federica Torrisi, Alessandro~Grilli, Tommaso~Catuogno, Mattia~Verducci, 
Elisabetta~Paladino, Luigi~Giannelli, Alessandro Sebastianelli
\thanks{C. Corradino and F. Torrisi are with the INGV-EO, Catania, Italy, e-mail: (claudia.corradino, federica.torrisi)@ingv.it,
T. Catuogno, A. Grilli and M. Verducci are with Thales Alenia Space Italia, e-mail: (tommaso.catuogno, alessandro.grilli-somministrato, mattia.verducci)@thalesaleniaspace.com. A. Sebastianelli is with Euro-Mediterranean Center on Climate Change, REMHI, Caserta, Italy, e-mail: alessandro.sebastianelli@cmcc.it. E. Paladino and L. Giannelli are with Dipartimento di Fisica e Astronomia “Ettore Majorana,” Università di Catania, Catania, Italy and with INFN, sezione di Catania, Catania, Italy, e-mail: (elisabetta.paladino, luigi.giannelli)@dfa.unict.it }
}

%
%

\markboth{}%
{Corradino \MakeLowercase{\textit{et al.}}: Hybrid Quantum Convolutional Neural Networks for Cross-Sensor Volcanic Thermal Activity Recognition at Global Scale}
%



\maketitle

\begin{abstract}
As Earth Observation (EO) enters the Big Data era, the exponential volume of daily satellite imagery poses significant computational and storage challenges for classical Deep Learning (DL) models. Moreover, current approaches often struggle to generalize across heterogeneous sensors and volcanic environments while requiring large labeled datasets and substantial computational resources. These limitations are particularly critical for emerging On-Board Processing (OBP) applications, where memory, computational power, and annotated data are inherently limited. 
This work proposes a Hybrid Quantum AlexNet architecture for cross-sensor recognition of volcanic thermal activity
at the global scale. The proposed model combines a classical convolutional backbone for high-level spatial features extraction with a parameterized quantum circuit (PQC) acting as a variational layer. By embedding high-level image representations into a high-dimensional Hilbert space, the quantum layer learns task-specific representations that enhance feature discrimination.
Experimental results demonstrate that the proposed hybrid quantum model learns more discriminative feature representations, leading to improved cross-sensor transferability and robustness across heterogeneous volcanic environments using fewer trainable parameters and reduced training data than its classical counterpart.
\end{abstract}


%
\IEEEpeerreviewmaketitle

\section{Introduction}
%
%
%
%
\IEEEPARstart{E}{arth} Observation (EO) has entered the Big Data era, with operational satellite missions generating massive volumes of multispectral imagery every day \cite{sudmanns2020big}. This unprecedented data availability enables continuous monitoring of natural hazards, including volcanic activity, but also demands efficient automatic analysis methods \cite{del2021artificial,tuia2024artificial}. Volcanic monitoring from space is particularly challenging because volcanic phenomena, such as intracrater thermal activity, lava flows, and lava lakes, exhibit high spatial and spectral variability worldwide, often under adverse atmospheric conditions \cite{harris2013thermal}. Deep Learning (DL), and in particular Convolutional Neural Networks (CNNs), has significantly improved volcanic scene classification from high spatial resolution satellite imagery \cite{corradino2023detection}.  However, CNN-based models are often sensor-dependent, requiring large labeled datasets and substantial computational resources for training. Their generalization capability may be compromised by the high variability of imaging conditions and volcanic environments encountered at the global scale. Recent transfer learning approaches have demonstrated promising results by reducing the dependence on large annotated datasets \cite{cariello2024}. Nevertheless, their effectiveness often deteriorates under significant domain shifts arising, for instance, from differences in both sensor characteristics (e.g., spatial and spectral resolution, radiometric response) and target volcanic environments (e.g., morphology, land cover, and surface manifestations of volcanic activity), thereby limiting their transferability across heterogeneous volcanic settings.
Hybrid Quantum Machine Learning (HQML) has emerged as a promising paradigm for representing and learning complex nonlinear relationships in the Hilbert space \cite{ miroszewski2023quantum}. By exploiting quantum superposition, entanglement, and high-dimensional quantum feature representations, hybrid quantum models may enhance the expressive power of the learning process, enabling richer representation learning and the extraction of highly nonlinear dependencies from heterogeneous data \cite{abbas2021}.
Hybrid Quantum-Convolutional Neural Networks (HQCNNs) in particular have recently emerged as a promising alternative for EO applications \cite{sebastianelli2025quantum, sebastianelli2021circuit}. By combining classical feature extraction with parameterized quantum circuits, HQCNNs exploit the expressive power of quantum states while remaining compatible with current Noisy Intermediate-Scale Quantum (NISQ) hardware \cite{kecceci2025accuracy, zaidenberg2021advantages}. We propose a HQCNN architecture for cross-sensor recognition of volcanic thermal activity from satellite imagery at global scale. This architecture embeds high-level visual features into a high-dimensional Hilbert space, where trainable parameterized rotation gates learn task-specific quantum representations, thereby enhancing class separability \cite{kecceci2025understanding}. This work, complementary to \cite{torrisi2026}, investigates, in a simulated quantum environment, whether quantum-enhanced neural networks can improve the efficiency and generalization capability of EO image classification, namely the ability to cope with domain shifts arising from differences in both sensor characteristics and volcanic environments.
The model is trained on Sentinel-2 (S2) MultiSpectral Instrument (MSI) imagery and evaluated on Landsat-9 (L9) Operational Land Imager (OLI) to assess  its cross-sensor generalization capability. 
Our study evaluates whether the proposed HQCNN can (i) achieve competitive classification performance with fewer trainable parameters than conventional CNNs, (ii) improve learning efficiency under limited labeled data, (iii) increase robustness to spectral variability, and geometric distortions, and (iv) enhance cross-domain generalization across heterogeneous volcanic environments and different EO sensors.
\section{Methodology}
\subsection{Data}
\begin{figure}[t]
\centering
\includegraphics[width=0.89\columnwidth,trim={0.1cm 1cm 0.1cm 1cm}]{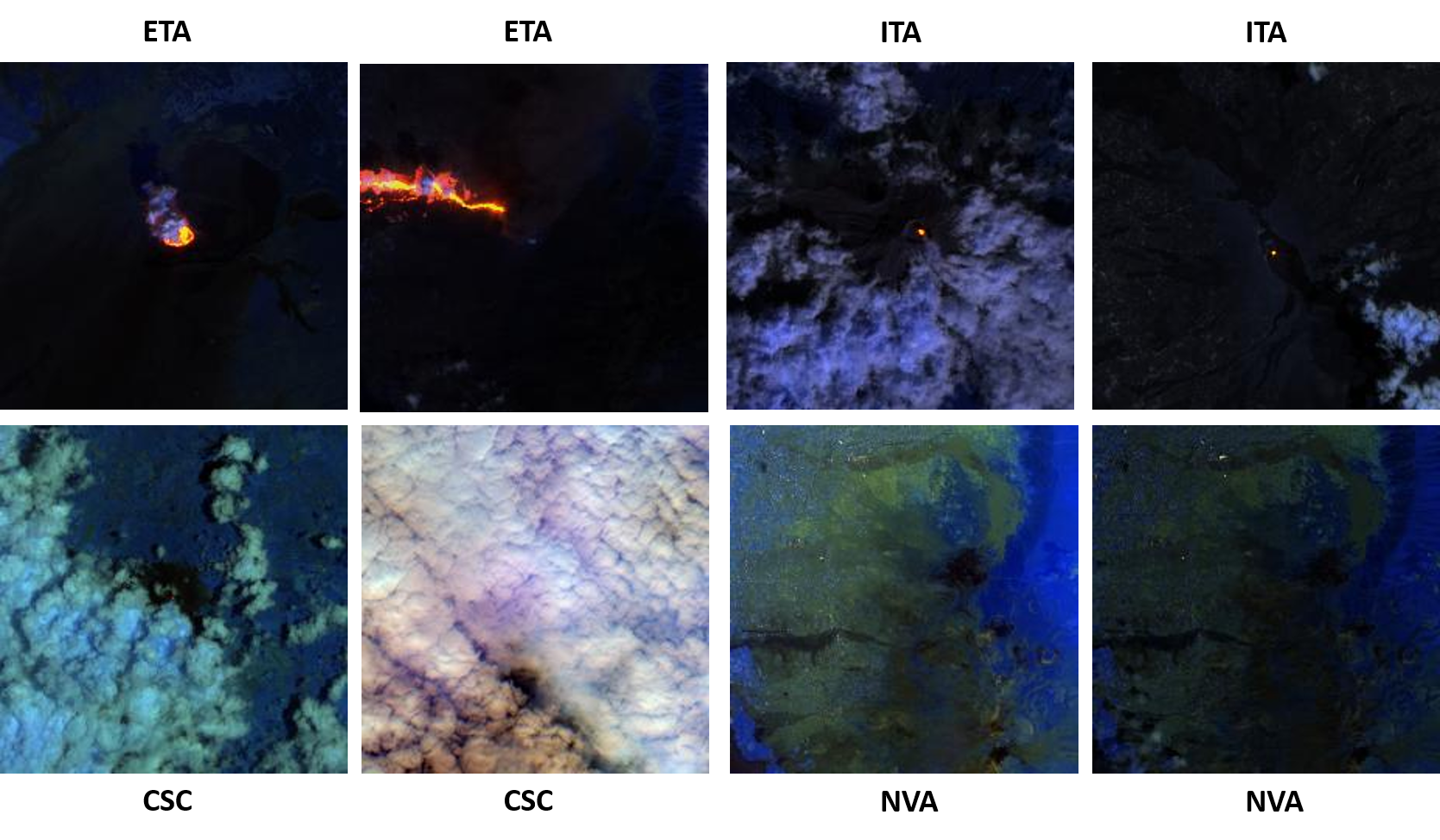}
\caption{Examples of S2 MSI images for the four dataset classes.}
\label{fig:EsempioC}
\end{figure}

The dataset is derived from the \textit{Copernicus Sentinel-2} mission, composed of the S2A and S2B satellites, each equipped with a MSI. Together, the two satellites acquire multispectral imagery in $13$ spectral bands from the visible to the ShortWave Infrared (SWIR), with a revisit time of approximately 5 days. For volcanic thermal monitoring, three bands are selected: NIR ($0.842$ $\mu$m), SWIR1($1.61$ $\mu$m), and SWIR2 ($2.20$ $\mu$m), owing to their sensitivity to high-temperature volcanic features and reduced atmospheric effects. Following the taxonomy proposed by Cariello \textit{et al.} \cite{cariello2024}, the dataset is labeled into four mutually exclusive classes (Fig.~\ref{fig:EsempioC}): {NVA} (No Volcanic Activity), {ITA} (Isolated Thermal Anomalies), {ETA} (Extended Thermal Anomalies), and {CSC} (Cloudy Sky Conditions). These classes represent the main volcanic observation scenarios encountered in optical satellite observations and provide a balanced benchmark. 
Images are preprocessed by generating a false-color composite (SWIR2, SWIR1, NIR), normalizing pixel values to the range $[0,1]$, and resizing them to $224 \times 224$ pixels. To evaluate cross-sensor generalization, the trained model is directly tested on L$9$ OLI imagery without additional fine-tuning. The corresponding spectral bands NIR ($0.865$ $\mu$m), SWIR1 ($1.608$ $\mu$m), and SWIR2 ($2.20$ $\mu$m) are combined using the same false-color composite adopted for S2. Despite differences in spatial resolution ($30$ m versus $10-20$ m) and spectral response, NIR and SWIR bands similarity enables a meaningful assessment of feature transferability across satellite missions. The balanced dataset includes $800$ images from several active volcanoes, namely Etna and Stromboli (Italy), Kilauea (USA), Popocat'epetl (Mexico), and Shiveluch (Russia). These volcanoes span different eruptive styles, including effusive, Strombolian, and lava dome extrusion/collapse activity, providing a diverse benchmark for evaluating the model robustness and generalization capability. 
\subsection{Hybrid Quantum Convolutional Neural Network}
\begin{figure*}[t]
\centering
\includegraphics[width=1\textwidth,trim=2cm 7cm 0cm 7cm,clip]{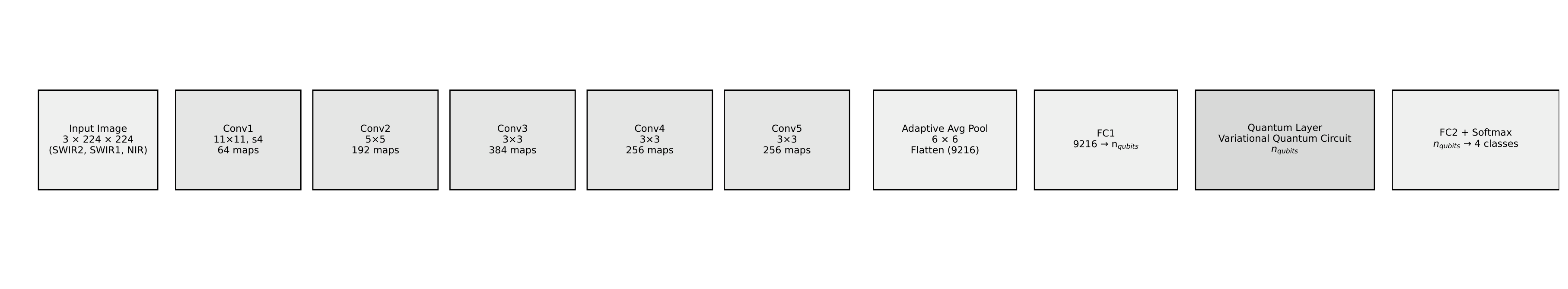}
\caption{Scheme of the proposed HQCNN: classical AlexNet backbone with a parameterized quantum
layer.}\label{fig:arch}
\end{figure*}

The proposed classifier is a HQCNN that combines a classical AlexNet backbone with a parameterized quantum layer (Fig.~\ref{fig:arch}) \cite{sebastianelli2021circuit}. This integrates QC into a DL framework by combining PyTorch, Qiskit, and Qiskit ML.
The architecture consists of four components, $(1)$ convolutional layers for hierarchical feature extraction, $(2)$  a fully connected projection layer to adapt the feature dimension to the quantum circuit input, $(3)$ a parameterized quantum layer, $(4)$ a final fully connected classification layer to map the quantum output to the target classes. 
The parameterized quantum layer operates on $n$ qubits and consists of two stages: a \textit{ZZFeatureMap} for data encoding and a variational \textit{RealAmplitudes} ansatz for trainable processing.The ZZFeatureMap encodes the classical feature vector into quantum states through Hadamard and $R_Z$ rotations, followed by entangling ZZ interactions that capture pairwise feature correlations. The trainable RealAmplitudes ansatz is composed of alternating $R_Y$ rotations and CNOT gates, enabling the circuit to exploit superposition and entanglement while learning task-specific representations. Specifically, using n qubits, the input data are embedded into a Hilbert space of dimension $2^{n}$, resulting in an exponentially growing feature space. This exponential scaling enhances the expressive capacity of the quantum model with respect to its classical counterpart.
Two HQCNN configurations employing $2$ and $4$ qubits were compared with classical AlexNet architectures whose fully connected layer ranged from $2$ to $4096$ neurons. The hybrid models are trained using the Adam optimizer with a learning rate of $10^{-3}$ and Cross-Entropy loss. The training-validation-test ratio is $70,15,15$. For comparison, a purely classical counterpart is implemented by replacing the quantum layer with a fully connected layer of varying size keeping the HQCNNs architecture and training settings \cite{sebastianelli2021circuit, sebastianelli2023quantum}. 


\section{Experimental Results}
\subsection{Classification Performance}
\begin{figure*}[!t]
    \centering

    \begin{minipage}[t]{0.485\textwidth}
        \centering
        \includegraphics[
            width=0.8\linewidth,
            trim={0.1cm 0.1cm 0.1cm 0.1cm},
            clip
        ]{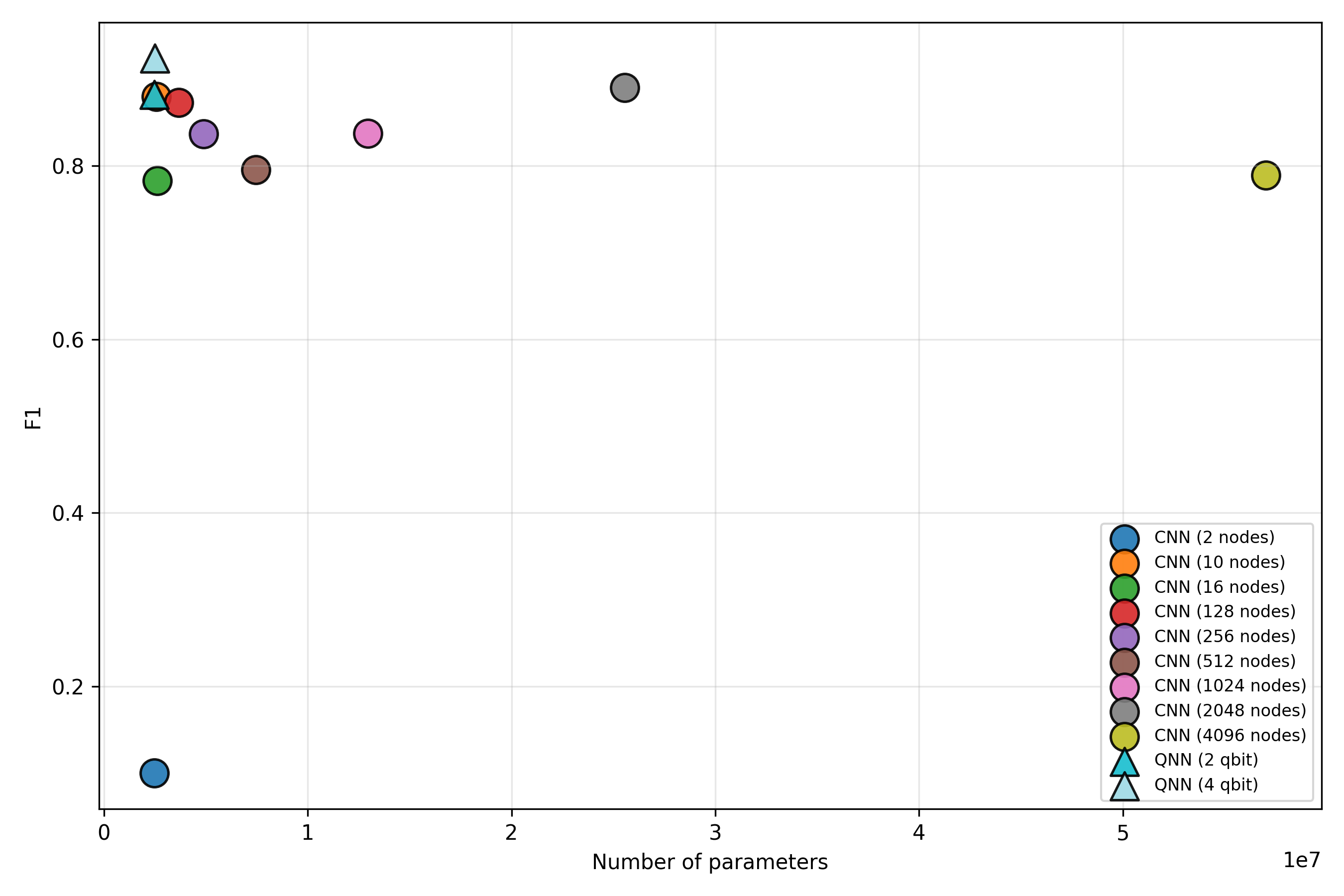}

        \vspace{1mm}
        \small (a) 
    \end{minipage}
    \hfill
    \begin{minipage}[t]{0.485\textwidth}
        \centering
        \includegraphics[
            width=0.81\linewidth,
            trim={0.1cm 0.11cm 0.1cm 0.1cm},
            clip
        ]{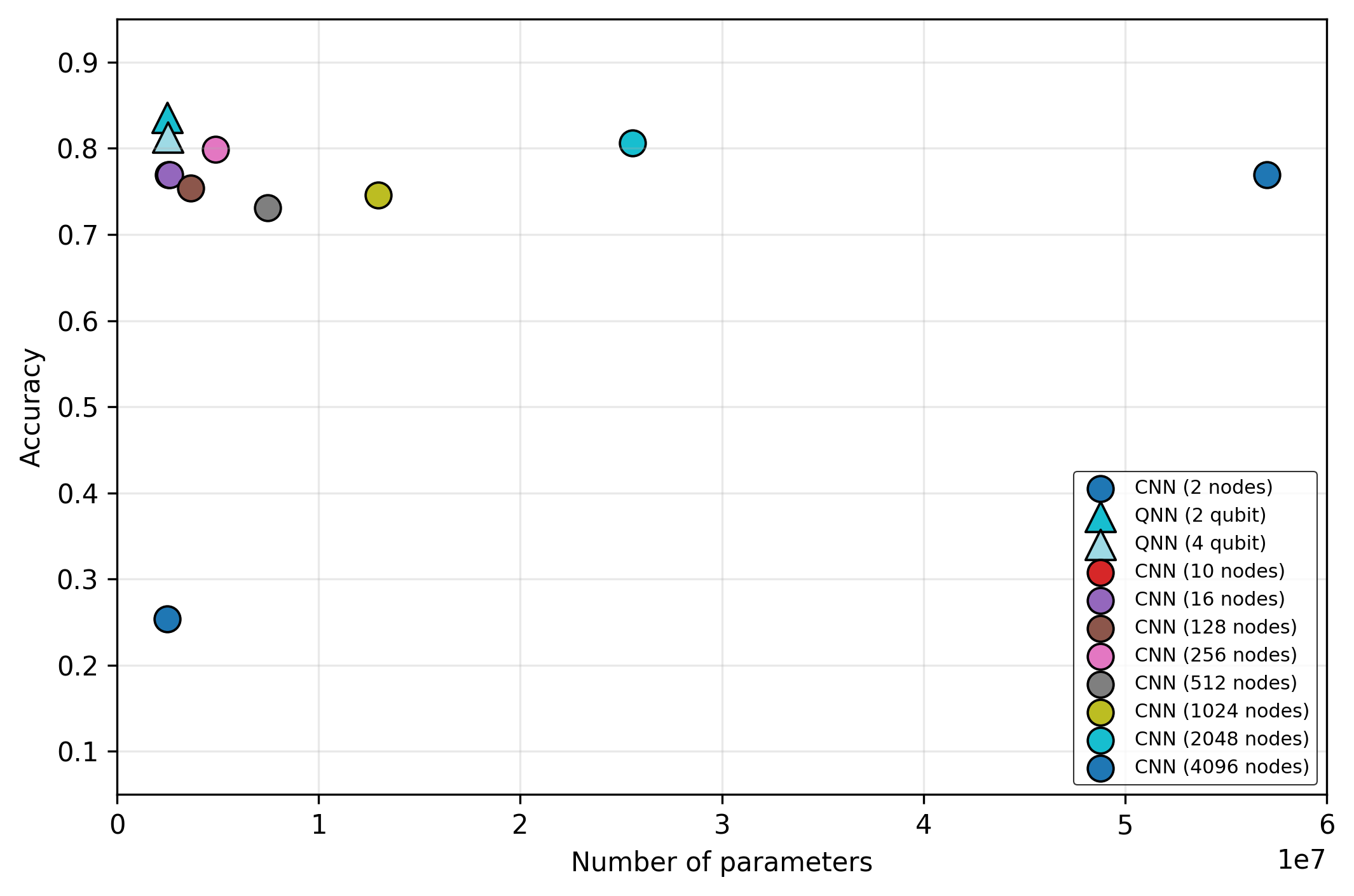}

        \vspace{1mm}
        \small (b)
    \end{minipage}

    \caption{Performances as a function of the number of trainable parameters for test scenes with similar (a) and different (b) orientations than the training data. }
    \label{fig:performance_vs_parameters}
\end{figure*}

Performance was evaluated using Accuracy, F1-score, Precision, and Recall. 
Table~\ref{tab:classification} summarizes the overall classification results. The proposed $4$-qubit HQCNN achieved the highest overall performance, reaching an Accuracy of $92.5$\% and an F1-score of $92$\% while requiring only $2.51$ M trainable parameters. The $2$-qubit HQCNN also achieved competitive performance (F1-score of $0.882$), outperforming classical models with a comparable parameter budget. Increasing the size of the classical fully connected layer did not systematically improve performance. While medium-sized networks ($10$-$2048$ nodes) achieved competitive results, architectures larger than $25$ M parameters exhibited a progressive degradation. In particular, the $4096$-node CNN reached an F1-score of only $0.79$\%, probably indicating overfitting caused by the limited amount of training data. Fig.~\ref{fig:performance_vs_parameters} (a) further highlights the relationship between predictive performance and model complexity. Beyond the global metrics, a class-wise analysis reveals that CSC and ETA are consistently the easiest class to discriminate, with several architectures achieving F$_1$-scores close to or above $0.90$\% and $0.87$\% respectively, whereas NVA and ITA represent the most challenging categories for most CNNs. In fact, while cloud and lava spectral and spatial structures are easy to learn, especially in cases where anomalies are very small and not easy to be recognized by the model, the variability of the volcanic scenarios at the global level makes the discrimination between the ITA and NVA not straightforward. The proposed HQCNNs substantially mitigate this limitation, achieving $82$\% (ITA) - $87$\% (NVA), and $89$\% (ITA) - $93$\% (NVA) for the $2$ and $4$ qubit configurations, respectively. Since the NVA and ITA classes are particularly relevant for operational volcanic monitoring, we report their average F$_1$-score as an indicator of the ability to simultaneously reduce false alarms during normal activity and detect small thermal anomalies. The highest values are obtained by the quantum counterpart indicating that quantum feature representations provide a more balanced classification across the different volcanic activity classes with lower model complexity than the best-performing CNNs.
\noindent
\setlength{\tabcolsep}{3pt}   
\begin{table*}[!t]
\centering
\caption{Classification performance of the investigated architectures.}
\label{tab:classification}
\small
\begin{tabular}{l r ccccc |ccccc}
\toprule
\textbf{Model} &
\textbf{Parameters} &
\begin{tabular}{c}\textbf{CSC}\\\textbf{F$_1$-Score}\end{tabular} &
\begin{tabular}{c}\textbf{ETA}\\\textbf{F$_1$-Score}\end{tabular} &
\begin{tabular}{c}\textbf{NVA}\\\textbf{F$_1$-Score}\end{tabular} &
\begin{tabular}{c}\textbf{ITA}\\\textbf{F$_1$-Score}\end{tabular} &
\begin{tabular}{c}\textbf{ITA+NVA}\\\textbf{F$_1$-Score}\end{tabular} &
\begin{tabular}{c}\textbf{Overall}\\\textbf{F$_1$-Score}\end{tabular} &
\begin{tabular}{c}\textbf{Overall}\\\textbf{Precision}\end{tabular} &
\begin{tabular}{c}\textbf{Overall}\\\textbf{Recall}\end{tabular} &
\begin{tabular}{c}\textbf{Overall}\\\textbf{Accuracy}\end{tabular} \\
\midrule
CNN (2 nodes)    & 2.488.148  & 0.00 & 0.00 & 0.00 & 0.40 & 0.20 & 0.10 & 0.25 & 0.06 & 0.25 \\
CNN (10 nodes)   & 2.562.020  & 0.94 & 0.92 & 0.82 & 0.85 & 0.83 & 0.88 & 0.88 & 0.88 & 0.88 \\
CNN (16 nodes)   & 2.617.508  & 0.94 & 0.82 & 0.68 & 0.70 & 0.69 & 0.78 & 0.78 & 0.84 & 0.78 \\
CNN (128 nodes)  & 3.666.500  & \textbf{0.98} & 0.90 & 0.82 & 0.79 & 0.80 & 0.87 & 0.88 & 0.87 & 0.88 \\
CNN (256 nodes)  & 4.896.068  & 0.92 & 0.88 & 0.77 & 0.78 & 0.78 & 0.84 & 0.84 & 0.84 & 0.84 \\
CNN (512 nodes)  & 7.453.508  & 0.87 & 0.90 & 0.68 & 0.74 & 0.71 & 0.80 & 0.80 & 0.82 & 0.80 \\
CNN (1024 nodes) & 12.961.604 & 0.91 & 0.89 & 0.76 & 0.80 & 0.78 & 0.84 & 0.84 & 0.84 & 0.84 \\
CNN (2048 nodes) & 25.550.660 & \textbf{0.98} & 0.92 & 0.85 & 0.81 & 0.83 & 0.89 & 0.89 & 0.89 & 0.89 \\
CNN (4096 nodes) & 57.020.228 & 0.85 & 0.87 & 0.61 & 0.83 & 0.72 & 0.79 & 0.80 & 0.81 & 0.80 \\
HQCNN (2 qubits)   & 2.488.150  & 0.94 & 0.90 & 0.87 & 0.82 & 0.85 & 0.88 & 0.88 & 0.88 & 0.88 \\
HQCNN (4 qubits)   & 2.506.600  & 0.92 & \textbf{0.95} & \textbf{0.89} & \textbf{0.94} & \textbf{0.91} & \textbf{0.92} & \textbf{0.93} & \textbf{0.93} & \textbf{0.93} \\
\bottomrule
\end{tabular}
\end{table*}
\setlength{\tabcolsep}{6pt}
\subsection{Efficiency}
\subsubsection{Parameter-to-Performance Efficiency}
To quantify the trade-off between predictive performance and model complexity, we define the  Parameter-to-Performance Efficiency,
\begin{equation}
\epsilon_{rel}=F_1\cdot\frac{P_{min}}{P_{model}},
\end{equation}
where $P_{min}$ is the minimum number of trainable parameters among the investigated architectures. The resulting efficiency ranking is reported in Table~\ref{tab:efficiency}. The proposed $4$-qubit HQCNN achieves the highest efficiency ($\epsilon_{rel}=0.917$), closely followed by the 2-qubit HQCNN ($0.882$). In contrast, the efficiency of classical CNNs rapidly decreases as the number of trainable parameters increases. For example, although the CNN with $4096$ nodes contains over twenty times more parameters than the HQCNNs, its $\epsilon_{rel}$ drops to $0.034$. These results suggests that quantum-enhanced feature processing provides a substantially higher information density per trainable parameter than increasing the size of the classical classifier.
\begin{table*}[!t]
\centering
\caption{Learning efficiency and cross-sensor generalization of the investigated architectures.}
\label{tab:efficiency}
\small
\begin{tabular}{l r cccc}
\toprule
\textbf{Model} &
\textbf{Parameters} &
$\boldsymbol{\epsilon_{rel}}$ &
\textbf{nAULC (\%)} &
\textbf{PW-nAULC} &
\textbf{Landsat-9 OLI F$_1$-Score} \\
\midrule
CNN (2 nodes)    & 2.488.148 & 0.10 & 34.13 & 11.07 & 0.41 \\
HQCNN (2 qubits)   & 2.488.150 & \textbf{0.88} & \textbf{78.91} & \textbf{56.44} & \textbf{0.88} \\
HQCNN (4 qubits)   & 2.506.600 & \textbf{0.92} & 74.24 & \textbf{51.42} & 0.60 \\
CNN (10 nodes)   & 2.562.020 & 0.86 & 71.53 & 47.57 & \textbf{0.80} \\
CNN (16 nodes)   & 2.617.508 & 0.74 & 71.07 & 51.35 & 0.65 \\
CNN (128 nodes)  & 3.666.500 & 0.60 & 72.48 & 44.72 & 0.76 \\
CNN (256 nodes)  & 4.896.068 & 0.43 & 70.19 & 44.07 & 0.79 \\
CNN (512 nodes)  & 7.453.508 & 0.27 & 69.29 & 44.02 & 0.75 \\
CNN (1024 nodes) & 12.961.604 & 0.16 & 73.87 & 49.01 & 0.78 \\
CNN (2048 nodes) & 25.550.660 & 0.09 & \textbf{75.64} & 50.12 & 0.77 \\
CNN (4096 nodes) & 57.020.228 & 0.03 & 72.70 & 41.55 & 0.80 \\

\bottomrule
\end{tabular}
\end{table*}

\subsubsection{Sample Efficiency}

\begin{figure}[!h]
\centering
\includegraphics[width=0.73\columnwidth,trim=0cm 0cm 0cm 0cm,clip]{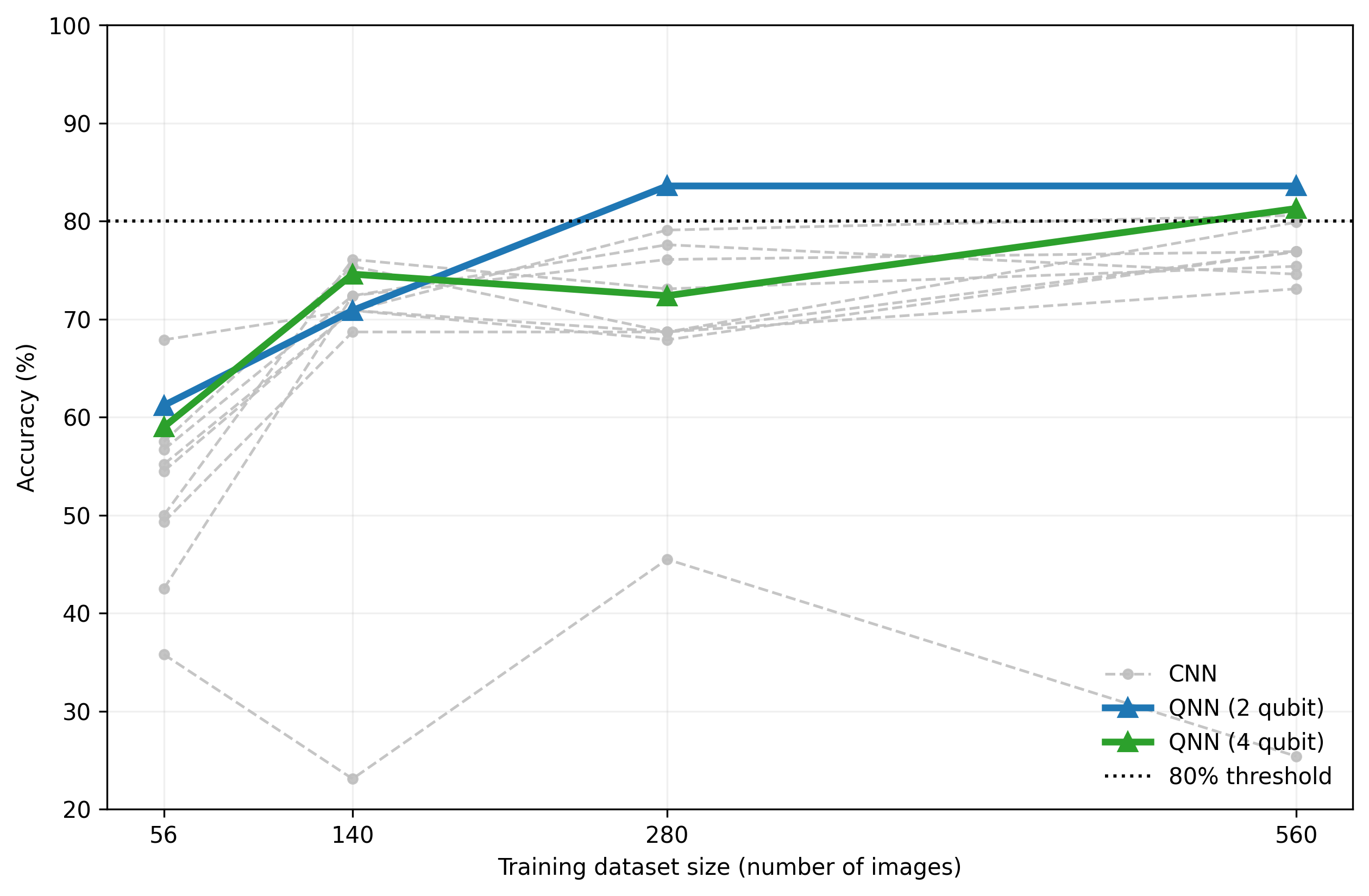}
\caption{Models accuracy for different training dataset sizes. }
\label{fig:accuracy_trends}
\end{figure}

To evaluate the sample efficiency of the proposed architectures, both the classical CNNs and the HQCNNs were trained using progressively smaller subsets of the original training dataset. Fig.~\ref{fig:accuracy_trends} shows the resulting learning curves. As expected, classification accuracy increases with the number of training samples for all architectures. However, the HQCNNs exhibit a steeper improvement, particularly in the low-data regime, indicating more effective exploitation of the available training samples. This trend is especially evident for the $2$-qubit HQCNN, which becomes the best-performing architecture from approximately $280$ training samples onward. Using an accuracy threshold of $80\%$ as reference, both HQCNNs reach this level with substantially fewer samples than the corresponding CNNs, highlighting their superior sample efficiency.

To quantify this behavior, we computed the normalized Area Under the Learning Curve (nAULC),
\begin{equation}
\mathrm{nAULC}=
\frac{1}{N_{\max}-N_{\min}}
\sum_{i=1}^{K-1}
\frac{A_i+A_{i+1}}{2}
\left(N_{i+1}-N_i\right),
\label{eq:naulc}
\end{equation}

where $A_i$ denotes the classification accuracy obtained using $N_i$ training samples. The nAULC represents the average accuracy over the investigated training range and provides a global measure of sample efficiency. To incorporate both the learning trajectory and the achieved performance, we define the Performance-Weighted nAULC (PW-nAULC),

\begin{equation}
\mathrm{PW\text{-}nAULC}
=
\mathrm{nAULC}
\sqrt{\frac{A_{\min}A_{\max}}{100^2}},
\label{eq:pwnaulc}
\end{equation}

where $A_{\min}$ and $A_{\max}$ denote the minimum and maximum classification accuracies. Table~\ref{tab:efficiency} reports the corresponding results. The $2$-qubit HQCNN achieves the highest nAULC and PW-nAULC, confirming its superior sample efficiency. The $4$-qubit HQCNN ranks second in terms of PW-nAULC, outperforming all investigated CNN architectures while requiring only $2.51$~M trainable parameters, while its learning behavior is comparable to that of the CNN ($2048$ nodes), which requires more than ten times as many parameters ($25.55$~M). Overall, the proposed quantum architectures achieve higher learning efficiency while maintaining competitive classification performance, making them particularly suitable for EO applications where labeled data and computational resources are limited.
\subsection{Generalization}
\subsubsection{Morphologic  robustness}
We can assess the robustness of the architectures to  geometric/morphologic variability using images obtained from rotation augmentation as test dataset (Fig.~\ref{fig:performance_vs_parameters} (b)). Although all models experienced a reduction in classification accuracy owing to the increased spatial complexity, the HQCNNs consistently exhibited a smaller performance degradation than the classical CNNs.
This behavior suggests that the quantum feature representation is less sensitive to orientation changes and preserves more discriminative information under geometric transformations. Overall, the results indicate that HQCNNs provide a favorable combination of classification accuracy, parameter efficiency, robustness to limited training data, and cross-sensor generalization, making them a promising solution for operational volcanic monitoring.

%
\subsubsection{Cross-Sensor Evaluation}
\begin{figure}[!t]
\centering
\includegraphics[width=1\columnwidth]{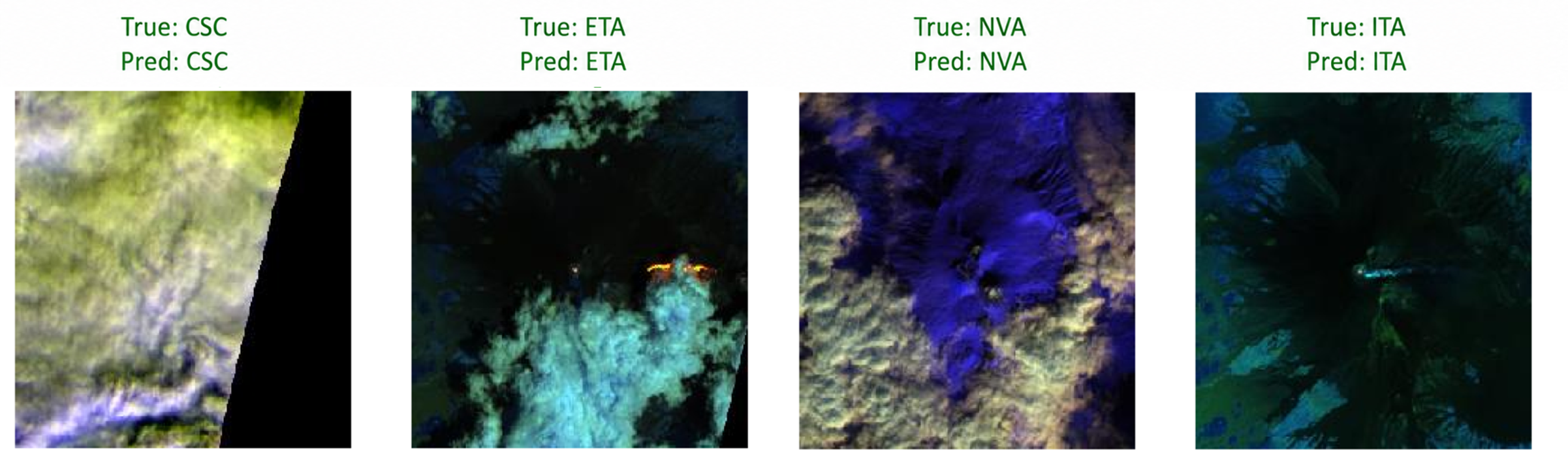}
\caption{L9 OLI classifications obtained with the proposed 2-qubit HQCNN. }
\label{fig:qcnn_examples}
\end{figure}

To assess the transferability of the learned representations, all models trained on S2 MSI imagery were directly evaluated on L9 OLI images without additional fine tuning. The results, reported in Table~\ref{tab:efficiency}, demonstrate that the proposed 2-qubit HQCNN achieves the highest overall F1-score ($0.88$), outperforming all classical baselines. 
This confirms that the learned quantum representations generalize effectively across sensors operating in comparable NIR and SWIR spectral ranges (Fig.~\ref{fig:qcnn_examples}).
However, the $4$-qubit HQCNN exhibited lower performance, suggesting that the larger quantum circuit learned representations that were more specialized to the S2 MSI training data and therefore less transferable to L9 OLI imagery. These results indicate that, in cross-sensor scenarios, a more compact quantum representation may provide better generalization than a higher-capacity quantum model.
\section{Conclusions and Future Perspectives}
This work demonstrates the potential HQCNNs for global volcanic thermal monitoring through the integration of a variational quantum layer within an AlexNet-based architecture. Experimental results show that the proposed quantum-enhanced models: (a) achieve competitive and, in several cases, superior performance compared with CNNs while using a comparable or significantly lower number of trainable parameters, and (b) provide an excellent trade-off among generalization capabilities, efficiency, and predictive
accuracy. In particular, while the $4$-qubit HQCNN achieved the highest overall classification performance, the $2$-qubit configuration provided the best trade-off solution.
This behavior is likely attributable to mild overfitting in the 4-qubit HQCNN, whose increased representational capacity may be excessive for the size and complexity of the available dataset. Following this evidence, two main conclusions can be drawn. Firstly, quantum architectures exhibit superior \emph{efficiency}. In terms of parameter-to-performance efficiency, they achieve competitive classification performance requiring only a small number of trainable quantum parameters. Unlike classical CNNs, whose performance improvements often rely on increasingly larger fully connected layers, the proposed architectures maintain high classification accuracy with only a few qubits, thereby reducing the risk of overfitting. Moreover, the proposed models demonstrate remarkable \emph{sample efficiency}, consistently outperform classical counterparts when trained with progressively smaller subsets of the dataset. This characteristic is particularly relevant for EO applications, where labeled volcanic data are often limited or highly imbalanced. Secondly, the proposed architecture exhibits strong \emph{generalization capability}. 
Robustness experiments show that HQCNNs experience a smaller performance degradation than classical CNNs, suggesting that the learned quantum representations are less sensitive to geometric variability. Furthermore, cross-sensor experiments performed on L9 OLI imagery without additional fine-tuning confirm that the learned representations are transferable across sensors operating in similar spectral ranges. Overall, these results indicate that HQCNNs provide a reliable and efficient alternative to conventional CNNs for cross-sensor EO image classification, offering enhanced parameter and sample efficiency, improved, and superior generalization across heterogeneous volcanic environments. Rather than relying on increasingly larger classical networks, the quantum-enhanced feature processing and the higher-dimensional representation provided by the quantum circuit improve the classification performance while maintaining a compact overall architecture thereby achieving greater parameter efficiency. Although the experiments were conducted in a simulated quantum environment, the competitive performance achieved with only 2 and 4 qubits suggests that meaningful quantum enhancement can already be obtained without requiring large-scale fault-tolerant quantum processors. Since only a small quantum register is required, the proposed approach is compatible with emerging compact NISQ devices, making future onboard deployment on EO satellite platforms increasingly feasible. Future work will investigate the impact of noise and limited coherence on classification performance, with the aim of running the quantum layer of the HQCNN on real, currently available, quantum hardware. Furthermore the role of quantum correlations in the learning process will be assessed by quantifying the amount of entanglement generated during inference. Quantum feature extraction through Quanvolutional NN will be investigated, enabling quantum processing within the convolutional stage. 
\section*{Acknowledgment}
C.C. thanks the Space It Up project funded by ASI and MUR under contract n. 2024-5-E.0 CUP n. I53D24000060005. L.G. and E.P. thank the PNRR MUR project PE0000023-NQSTI and Università degli Studi di Catania, project TCMQI PIACERI 2024/2026. E.P. thanks the COST Action CA21144 SuperQumap.

\ifCLASSOPTIONcaptionsoff
  \newpage
\fi



\bibliographystyle{IEEEtran}
%

%





\end{document}